\documentstyle[aps,prb,epsfig]{revtex}
\begin{document}
\wideabs{
\title{Magnetic Properties of HTSC with Weak Interlayer Coupling.}
\author{Gregory M. Braverman}
\address{Max-Planck-Institut f\"ur Kernphysik Heidelberg, Germany.}
\maketitle
\begin{abstract}
Magnetic properties of layered high temperature
superconductors with a weak interlayer coupling in the region of
critical fluctuations are investigated in framework of the
Ginzburg--Landau approach. The sample magnetization is calculated  
perturbatively to
the second order in the interlayer coupling constant. The first order
correction can be incorporated into the non-interacting expression for the
magnetization with the corresponding shift of the critical temperature.
Only the second order contribution modifies thermodynamics, leading to
violation of the 2D scaling law and to disappearance of the
magnetization crossing point. Magnetic field suppresses the interlayer
interaction, and at sufficiently high values of the applied field 2D
scaling is recovered.
\end{abstract}
}

\section{Introduction.}
Scaling properties of layered high temperature superconductors (HTSC)
around the mean--filed transition line $H_{c2}(T)$ has
been under intensive experimental\cite{welp,li} and  
theoretical\cite{brezin,ulla,ikeda,thouless,larkin,tesb,tesandr,pierson}
study
during the
last
decade. The layered Y\-Ba$_2$\-Cu$_3$\-O$_{7-\delta}$ compounds exhibit 3D 
properties\cite{welp}. All thermodynamic and transport 
quantities obey the 3D scaling law as a function of the scaling variable
$B(T-T_{c2}(H))/(TH)^{2/3}$, when the applied magnetic field is
perpendicular to the layers plane. In the case of
Bi$_2$\-Sr$_2$\-Ca$_2$\-Cu$_3$\-O$_{10}$ experimental results\cite{li}
show
2D scaling behaviour with respect to the 2D scaling variable
$A(T-T_{c2}(H))/(TH)^{1/2}$. The fact that the phenomenon is observed in
the vicinity of the mean--field upper critical field indicates that the
problem can be studied in the framework of the Ginzburg--Landau (GL)
theory with the order parameter projected onto the space of the lowest
Landau levels (LLL). First calculations of the
free energy scaling function were done
perturbatively\cite{brezin,ulla,ikeda,thouless,larkin}. The
non-perturbative
approach for pure 2D systems was developed by Tesanovi\'c {\it et.
al.}\cite{tesb}. It was shown that around the mean--field
transition line only the fluctuations of the total amplitude of the order
parameter play important role in the superconductor thermodynamics. The
remaining part, fluctuations in the position of the vortices
enters through the dimensionless Abrikosov\cite{abrikosov} geometric
factor $\beta_A$. In the most realistic cases this quantity weakly depends
on the vortex configuration and can be taken as a constant from the very
beginning. This approach was generalized for the layered 2D and 3D systems  
by Tesanovi\'c and Andreev\cite{tesandr}. They showed
that the system dimensionality depends on the strength of the interlayer
coupling constant. When coupling is absent, the system becomes
pure two dimensional. In the limit of strong interlayer coupling, when
the superconducting correlation length $\xi_c$ along the $c$-axis
direction becomes much larger then the effective interlayer separation $s$
the 3D description becomes appropriate. In the intermediate case $\xi_c <
s$ the system becomes quasi two--dimensional for which scaling is
impossible.\\

In this Paper we study quasi two dimensional layered superconductor
in the region of the critical fluctuations. In this region we use GL-LLL
description, which allows us to build a solvable model. The interlayer
coupling is assumed to be small and therefore can be taken into account
perturbatively. Within this approach we calculated the sample
magnetization to the second order in the interlayer coupling constant. If
this constant is extremely small one can consider only the first order
correction. We show that because of the nearest neighboring coupling this
correction leads to shift of the critical temperature, preserving the
form of usual 2D scaling function of the magnetization. Only the next
order correction destroys scaling and leads to
disappearance of the magnetization crossing point. The effective
interlayer coupling constant turns out to be proportional to the
square root of the inverse
applied field, which results in effective suppression of the interlayer
interaction in high field region and, therefore, leads to recovering of
the 2D scaling law.

\section{The Model.}
Consider a layered type II superconductor in the region of critical
fluctuations around its mean-field transition line $T_{c2}(H)$ (or
$H_{c2}(T)$). The applied magnetic field ${\bf H}$ is assumed to be normal
to the layers plane: ${\bf H}\parallel\hat{c}$. Then the superconductor 
thermodynamics at the temperature $T$ can be described by the following
partition function:
\begin{equation}
{\cal Z}\propto\int{\cal D}[\Psi]\int{\cal D}[{\bf A}]
\exp\left\{
-\frac{{\cal F}[\Psi,{\bf A}]}{k_BT}
\right\},
\label{partf.eq}
\end{equation}
where ${\cal F}[\Psi,{\bf A}]$ is the GL free energy
functional of the layered system with a nearest neighboring
Josephson coupling between pancake vortices given by
\begin{eqnarray}
{\cal F}[\Psi,{\bf A}]=
s\sum_n\int d^2{\bf r}\left\{
\alpha_0|\Psi_n|^2+\frac{\beta}{2}|\Psi_n|^4+
\nonumber\right.\\\left.
\gamma_{ab}|\partial_n\Psi_n|^2+
\gamma_c|\Psi_n-\Psi_{n+1}|^2+\frac{({\bf H}-{\bf B}_n)^2}{8\pi}
\right\}
\label{fren.eq}
\end{eqnarray}
and $k_B$ is the Boltzmann constant. The quantity $s$ is an effective
interlayer spacing, $\alpha_0=a(T-T_{c0})$ and
$\beta=const$ are the first and the second GL coefficients
correspondingly. The third GL coefficient $\gamma$ is assumed to be
anisotropic, where the quantities $\gamma_{ab}$ and $\gamma_c$ define its
value in the layer plane and $\hat{c}$ direction correspondingly. In what
follows, we refer to the quantity $\gamma_c$ as an interlayer coupling
constant. The
quantity $\Psi_n({\bf r})$ is the order parameter of the $n$th layer and
${\bf B}_n({\bf r})\parallel\hat{c}$ is the magnetic induction induced in
the $n$th layer of the superconductor. Two-dimensional gauge invariant
gradient $\partial_n$ is defined as:
$$
\partial_n=-i\hbar\frac{\partial}{\partial{\bf r}}-
\frac{2e}{c}{\bf A}_n({\bf r}),
$$
where ${\bf A}_n({\bf r})=\nabla\times {\bf B}_n({\bf r})$. The sample
magnetization is given by
\begin{equation}
{\bf M}=\frac{1}{4\pi N_L}\sum_n{
\frac{
\int{\cal D}[\Psi,{\bf A}]({\bf B}_n-{\bf H})
\exp\left\{
-\displaystyle{\frac{{\cal F}}{k_BT}}
\right\}
}
{
\int{\cal D}[\Psi,{\bf A}]
\exp\left\{
-\displaystyle{\frac{{\cal F}}{k_BT}}
\right\}
}
},
\label{m.eq}
\end{equation}
where $N_L$ is the number of layers. In the weak coupling regime, it is
convenient to introduce renormalized critical temperature:
\begin{equation}
T_c= T_{c0}-2\frac{\gamma_c}{a}\equiv T_{c0}-\delta T_c.
\label{tshift.eq}
\end{equation}
This modifies the expression (\ref{fren.eq}) for the free energy, which
now reads as 
\begin{eqnarray}
{\cal F}[\Psi,{\bf A}]=
s\sum_n\int d^2{\bf r}\left\{
\alpha|\Psi_n|^2+\frac{\beta}{2}|\Psi_n|^4+
\nonumber\right.\\\left.
\gamma_{ab}|\partial_n\Psi_n|^2-
\gamma_c\Psi^*_n(\Psi_{n-1}+\Psi_{n+1})+
\frac{({\bf H}-{\bf B}_n)^2}{8\pi}
\right\},
\label{fren1.eq}
\end{eqnarray}
where $\alpha=a(T-T_c)$. Thus, in order to solve the problem one has to
calculate extremely complicated integrals over the order parameter $\Psi$
and the vector potential $\bf A$ appearing in the expressions
(\ref{partf.eq}) and (\ref{m.eq}) for the partition function and the
magnetization. However, there are number of simplifications,
which can be done.\\

In the limit of large values of the GL parameter $\kappa$ ($\kappa\sim
100$ for the most HTSC) one can
neglect fluctuations of the magnetic induction ${\bf B}_n$ \cite{tes}.
Then we minimize the last expression (\ref{fren1.eq}) for the GL free
energy
with respect to the vector potential ${\bf A}_n$ as it is done in the case
of conventional superconductor. This leads to the set of decoupled (with
respect to the layer index) GL
equations for the vector potential ${\bf A}_n$ and the order parameter
$\Psi_n({\bf r})$. 
In general, the order parameter can be expanded over the electron
eigenfunctions of the
Landau levels in the applied magnetic field $H$. However, close to
the mean-field transition line $H_{c2}(T)=-\alpha(T)c/(2\hbar
e\gamma_{ac})$ one can restrict oneself to
the zeroth Landau level only. This approximation can be
used, at least, if $H>1/3H_{c2}(T)$, until the first Landau level becomes
important.
Recently, it was shown \cite{kita,liRos} that the lowest Landau level
(LLL) approximation works good even if $H\ll H_{c2}(T)$ for $\kappa\gg 1$.
In the LLL approximation the equations, described above, can
be solved analytically \cite{abrikosov}. After substituting these
solutions into the expression (\ref{fren1.eq}) for
the free energy, we finally obtain:
\begin{eqnarray}
{\cal F}=sS\sum_n\left\{
\alpha\left(1-\frac{H}{H_{c2}}\right)\overline{|\Psi_n|^2}
+\frac{\beta}{2}\overline{|\Psi_n|^4}-
\right.\nonumber\\\left.
\gamma_c\overline{\Psi^*_n(\Psi_{n-1}+\Psi_{n+1})}
\right\},
\label{fren3.eq}
\end{eqnarray}
where the bar means averaging over the layer area $S$.
In this case only the integrals over the order parameter $\Psi$ remain
in the expression for the partition function (\ref{partf.eq}):
$$
{\cal Z}\propto\int{\cal D}[\Psi]\exp\left[
-\frac{{\cal F}[\Psi]}{k_BT}
\right],
$$
where the quantity ${\cal F}[\Psi]$ is given now by (\ref{fren3.eq}).
In the LLL approximation the sample magnetization (\ref{m.eq}) can
be calculated using the following formula:
\begin{equation}
M=\frac{\beta H_{c2}}{8\pi\alpha\kappa^2}
N_L^{-1}\sum_n\frac{
\int{\cal D}[\Psi]\overline{|\Psi_n|^2}
\exp\left(-\displaystyle{\frac{{\cal F}}{k_BT}}\right)
}
{
\int{\cal D}[\Psi]
\exp\left(-\displaystyle{\frac{{\cal F}}{k_BT}}\right)
}.
\label{m1.eq}
\end{equation}
In order to proceed further, we
define the Abrikosov geometric factor for each layer separately: 
$\beta_A(n)\equiv\overline{|\Psi_n|^4}/
\left(\overline{|\Psi_n|^2}\right)^2$, and its
average $\beta_A=N_L^{-1}\sum\beta_A(n)$. Following the
refs. \cite{tesb,tes1} we
assume that the quantity $\beta_A(n)$ only slightly depends on the actual
vortex configuration and therefore we put $\beta_A(1) = \beta_A(2)= ... =
\beta_A$. Then we
replace the quantities $\overline{|\Psi_n|^4}$
by $\beta_A\left(\overline{|\Psi_n|^2}\right)^2$ in the expression
(\ref{fren3.eq}) for the free energy. With this replacement the model
becomes exactly solvable:
\begin{eqnarray}
{\cal Z}\propto\int{\cal D}[\Delta]
\exp\left\{
-N_v\sum_n\left[
x\overline{|\Delta_n|^2}+
\frac{1}{4}\left(\overline{|\Delta_n|^2}\right)^2
\right.\right.\nonumber\\\left.\left.
-\mu\overline{\Delta_n^*(\Delta_{n+1}+\Delta_{n-1})}
\right]
\right\},
\label{partf1.eq}
\end{eqnarray}
where $N_v=\Phi/\Phi_0$ is the number of vortices. The standard 2D scaling
variable $x$ is given by
$$
x=A\frac{T-T_{c2}(H)}{\sqrt{TH}},
$$
where $A=\sqrt{s\Phi_0/(16\pi\kappa^2\beta_Ak_B)}H_{c2}^\prime$ and
$H_{c2}^\prime=-dH_{c2}(T)/dT|_{T=T_c}$. The dimensionless coupling
constant $\mu$ is
\begin{equation}
\mu=A\frac{\delta T_c}{2\sqrt{TH}}
\label{mu.eq}
\end{equation}
and the dimensionless order parameter $\Delta_n$ reads as 
$$
|\Delta_n|^2=A\frac{2\beta_A\beta}{a\sqrt{TH}}|\Psi_n|^2.
$$
With these new variables the sample magnetization (\ref{m1.eq}) can be
expressed as
\begin{equation}
\frac{M}{\sqrt{HT}}=\frac{k_BA}{s\Phi H_{c2}^\prime}N_L^{-1}
\frac{d\ln{\cal Z}}{dx}.
\label{m2.eq}
\end{equation}

\section{Calculation of the Partition Function.}
In order to compute magnetization of the superconductor we, first,
have to calculate the partition function (\ref{partf1.eq}). This involves
evaluation of the integrals over the order parameter $\Delta$. The main
difficulty in such calculation is the quartic term appearing in the
exponent of the right hand side of the formula (\ref{partf1.eq}). In order
to decouple it, we introduce a set of additional integration variables
$\{\lambda_n\}$:
\begin{eqnarray}
{\cal Z}\propto\int{\cal D}[\Delta]\int_\Gamma{\cal D}[\lambda]
\exp\left\{
-N_v\sum_n\left[
\lambda_n^2+
\right.\right.\nonumber\\\left.\left.
(x+i\lambda_n)\overline{|\Delta_n|^2}-
\mu\overline{\Delta_n^*(\Delta_{n+1}+\Delta_{n-1})}
\right]
\right\},
\label{partf2.eq}
\end{eqnarray}
where 
$$
\int_\Gamma{\cal D}[\lambda]=\prod_n\int_{\Gamma_n}d\lambda_n .
$$ 
The contours $\Gamma_n$ are parallel to the real axis, standing on
some
distance from it, in order to insure convergence of the integrals over the
order parameter $\Delta$. The formula (\ref{partf2.eq}) is the result of
use of the simplified version of
the Hubbard-Stratonovich transformation, usually applied in the field
theory. As it was explained above, the order parameter $\Delta_n$ is the
linear combination of the electron eigenfunctions of the lowest Landau
level:
$$
\Delta_n({\bf r})=\sum_{k=0}^{N_v}C_{nk}L_k({\bf r}),
$$
and
$$
L_k({\bf r})=\frac{1}{\sqrt{k!}}\left(\frac{r}{l}\right)^k
\exp\left\{
-ik\vartheta-\frac{r^2}{2l^2}
\right\},
$$
where $l$ is the magnetic length corresponding to the charge $2e$.
Then the meaning of the integration over the order parameter becomes
clear: 
$$
{\cal D}[\Delta]\propto\prod_{n,k}\int dC^*_{nk}dC_{nk}.
$$ 
The integrals over these expansion coefficients in (\ref{partf2.eq}) are 
of
the generalized gaussian type and can be evaluated analytically. As a
result, we
obtain:
\begin{equation}
{\cal Z}\propto\int_\Gamma{\cal D}[\lambda]\exp\left\{-N_v{\cal
L}\right\}.
\label{partf3.eq}
\end{equation}
The action $\cal L$ is given by
$$
{\cal L}=\textrm{tr}\left[
\hat{\lambda}^2
+\ln\left(
x\hat{\textrm{I}}+i\hat{\lambda}-\mu\hat{\gamma}
\right)
\right],
$$
where $\hat{\lambda}$ is diagonal matrix, consisting from
the elements $\lambda_n$. The matrix $\hat{\gamma}$
is the real symmetric matrix, arising as a result of the interlayer
coupling:
\begin{equation}
\gamma_{mn}=\delta_{m,n+1}+\delta_{m,n-1}.
\label{gamma.eq}
\end{equation}
The integrals over the expansion coefficients $C_{nk}$ converge, if along
the integration contours $\Gamma_n$ the following inequality is satisfied:
\begin{equation}
\textrm{Re}(f_n)>0,
\label{apc.eq}
\end{equation}
where $f_n$ are eigenvalues of the complex symmetric matrix
$x\hat{\textrm{I}}+i\hat{\lambda}-\mu\hat{\gamma}$. 
In the thermodynamic limit $N_v\rightarrow\infty$ the integrals over
$\lambda_n$ in the partition function (\ref{partf3.eq}) can be evaluated
within saddle point approximation. It will be shown below that due to
particular properties of the saddle-point manifold the condition
(\ref{apc.eq}) can
be satisfied in the weak-coupling regime in rather large range of the
scaling variable $x$.\\

The saddle point equation is found from the condition $\delta{\cal L}=0$
and reads as
\begin{equation}
\textrm{tr}\left\{\left[
2\hat{\lambda}+
i\left(x\hat{\textrm{I}}+i\hat{\lambda}-\mu\hat{\gamma}\right)^{-1}
\right]\delta\hat{\lambda}\right\}
=0.
\label{sp.eq}
\end{equation}
The position of the saddle point depends now on the temperature $T$ and
the 
applied field $H$ via two parameters $x(H,T)$ and
$\mu(H,T)$. In this case, 2D scaling of the
magnetization (\ref{m3.eq}) becomes impossible, as it was predicted in the
ref. \cite{tesandr}. If the
coupling is weak $2\mu< |x+i\lambda|$ (this inequality is similar to
Tesanovi\'c-Andreev criterion \cite{tesandr} for quasi 2D systems), the
second term in the left
hand side of the equation (\ref{sp.eq}) can be expanded in the powers of
$\mu$. Then the saddle point equation (\ref{sp.eq}) can be rewritten as
follows:
\begin{eqnarray}
&2\lambda_n+\displaystyle{\frac{i}{x+i\lambda_n}}+&\nonumber\\
&\displaystyle{\frac{i\mu^2}{(x+i\lambda_n)^2}}
\displaystyle{\frac{2x+i(\lambda_{n-1}+\lambda_{n+1})}
{(x+i\lambda_{n-1})(x+i\lambda_{n+1})}}
+o(\mu^4)=0.&
\label{sp1.1.eq}
\end{eqnarray}
The terms of the order of $\mu$ and $\mu^3$ drop out in the right hand
side of the
last equation, since $\textrm{tr}(\hat{\gamma}^{2n+1})=0$. The structure
of the saddle point equation is such that, at least, up to the
second  order in
$\mu$ the saddle point solution for $\hat{\lambda}$ is proportional to the
unit matrix, namely
$$
\lambda_1=\lambda_2= ... \equiv\lambda.
$$
Then the equation (\ref{sp1.1.eq}) can be rewritten in the following
simple form:
\begin{equation}
2\lambda+\frac{i}{x+i\lambda}+
\frac{2i\mu^2}{(x+i\lambda)^3}
+o(\mu^4)=0.
\label{sp2.eq}
\end{equation}
As soon as the saddle point solution is found,
the sample
magnetization (\ref{m2.eq}) can be calculated using the following simple
relation:
\begin{equation}
\frac{M}{\sqrt{HT}}=-\frac{2Ak_B}{s\Phi_0H_{c2}^\prime}
i\lambda.
\label{m3.eq}
\end{equation}
From the last equation we conclude that the physically meaningful solution
for $\lambda$ lies on the imaginary axis. Further, using the
Rayleigh-Ritz theorem \cite{Lut}, one can show that both matrices
$2\hat{\textrm{I}}-\hat{\gamma}$ and $2\hat{\textrm{I}}+\hat{\gamma}$
are positive definite. Then the eigenvalues of the matrix $\hat{\gamma}$
belong to the interval $[-2,2]$. In this case, the conditions
(\ref{apc.eq}) for 
convergence of the integrals over expansion coefficients $C_{kn}$ in the
partition function can be written as:
\begin{equation}
2\mu<|x+i\lambda|,
\label{app.eq}
\end{equation}
which is automatically satisfied in the weak coupling regime.

\section{Magnetic Properties.}
\label{magprop.seq}
In the previous section we derived the saddle point equation
(\ref{sp2.eq}) for the
layered superconductor under assumption that the interlayer coupling
constant $\mu$ is small. The sample magnetization $M(H,T)$ is proportional
to the saddle point solution and is given by the formula (\ref{m3.eq}).
\begin{figure}
\begin{center}
\epsfxsize=8.5cm 
\epsfbox{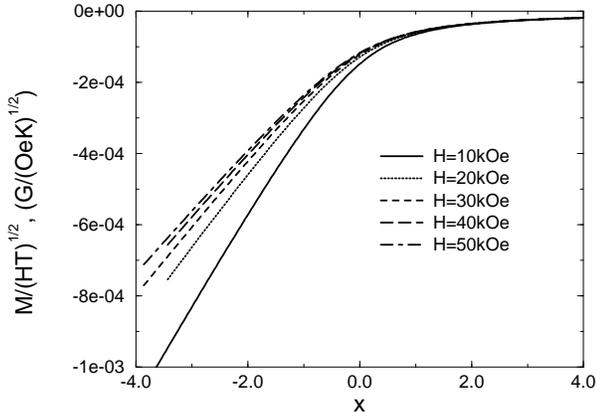}
\end{center}
\caption{The quantity $M/\sqrt{HT}$ as a function of the 2D scaling
variable $x$ is plotted for $\delta T_c=1K$.}
\label{fig1}
\end{figure}
In the "zeroth" approximation only one of the two saddle point solutions
of the eq. (\ref{sp2.eq}) can be
reached by allowed deformation of the integration contour:
$$
\lambda_0(x)=\frac{i}{2}(x-\sqrt{x^2+2}).
$$
Then the magnetization preserves the 2D scaling law and is given
by the same formula as in the noninteracting case \cite{tesb}:
\begin{equation}
\frac{M_0}{\sqrt{HT}}=\frac{Ak_B}{s\Phi_0H_{c2}^\prime}
(x-\sqrt{x^2+2}).
\label{m0.eq}
\end{equation}
Actually, this "zeroth" order result includes the first order correction
in $\mu$ by means of the critical temperature shift (\ref{tshift.eq}). The
second order correction to the saddle point solution is calculated as a
small perturbation:
\begin{equation}
\lambda(x,\mu)=\lambda_0(x)
\left(
1+4\mu^2\frac{i\lambda_0(x)}{\sqrt{x^2+2}}
\right).
\label{sps.eq}
\end{equation}
Then the sample magnetization is given by
\begin{equation}
M=M_0\left(
1-2\mu^2\frac{x-\sqrt{x^2+2}}{\sqrt{x^2+2}}
\right),
\label{m4.eq}
\end{equation}
where $M_0(H,T)$ is the magnetization of the "decoupled" sample given by
the equation (\ref{m0.eq}).
Using the last formula, we plotted the quantity $M/\sqrt{HT}$ as a
function of 2D scaling variable $x$ (see the figure \ref{fig1}) for
five different values of the applied field $H$ between $10$ and
$50kOe$. The interlayer coupling constant is chosen to be small, such that
$\delta T_c=1K$(see the eq. \ref{tshift.eq} for definition of $\delta
T_c$). We used $T_c=111K$, $\kappa=100$,
$H_{c2}^\prime=40kOeK^{-1}$
and
$s=2nm$, which are of the order of the typical experimental
parameters \cite{li}.
The whole range of the scaling variable $x$ in the fig. \ref{fig1}
satisfies the
applicability condition of the theory (\ref{app.eq}). As it was expected,
the 2D scaling is destroyed, due to the dimensionless coupling
constant $\mu(T,H)$
appearing in the right hand side of the expression (\ref{m4.eq}) for the
magnetization.\\ 
\begin{figure}
\begin{center}
\epsfxsize=8.5cm
\epsfbox{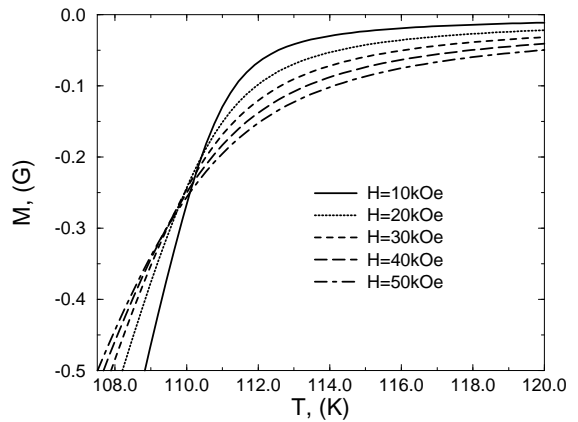}
\end{center}
\caption{The sample magnetization $M$ as a function of the temperature $T$ 
is plotted for $\delta T_c=1K$.}
\label{fig2}
\end{figure}

\begin{figure}
\begin{center}
\epsfxsize=8.5cm
\epsfbox{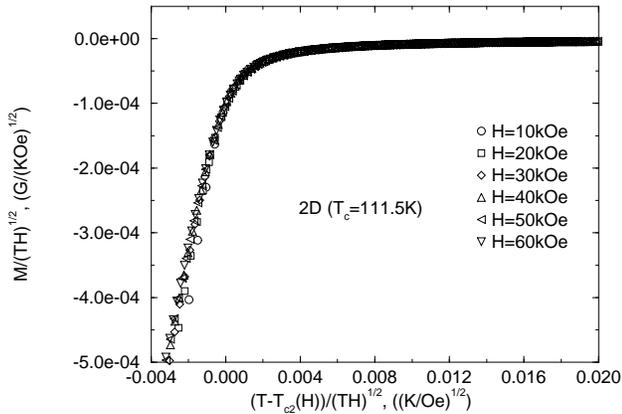}
\end{center}
\caption{2D scaling of the magnetization data, taken from the fig.
\ref{fig2}.}
\label{fig3}
\end{figure}

The plot of the magnetization as a function of temperature can be
found in the fig. \ref{fig2}, in which we used the same set of the
phenomenological parameters, as for the fig. \ref{fig1}.
Like in the previous case, the whole temperature range in this
figure
satisfies the applicability condition (\ref{app.eq}).
Form this figure follows that even account of the small interlayer
coupling destroys
the magnetization crossing point, which is the property of the scaling
form (\ref{m0.eq}) (see ref. \cite{tesb}). Actually, the crossing point is
noticeably destroyed only in the low--field region. The higher the
applied field, the better the crossing point is pronounced. This is the
consequence of the specific form of the effective coupling constant $\mu$
(see eq. (\ref{mu.eq})). The interaction correction to the magnetization
in
the formula (\ref{m4.eq}) is proportional to the square of this
constant and therefore is proportional to $H^{-1}$. Then, at
the large values of the applied field interlayer interaction is
effectively suppressed. Indeed, one can treat the magnetization data
plotted in the figure \ref{fig2} as if they were obtained from the
experiment. Then, we try to fit them to 2D scaling form variating
phenomenological parameters $T_c$ and $H_{c2}^\prime$, as it is usually
done in experiments. The fitting results are given in figure \ref{fig3}. 
The best fit is obtained for $T_c\approx 111.5K$. Like in the
experiment\cite{li}, the scaling results are insensitive to 
the value of the phenomenological parameter $\mu_0H_{c2}^\prime$ in
relatively large range of its values around $4TK^{-1}$. It can be
observed from the figure \ref{fig3} that
the scaling is satisfactory good for fields larger than $30kOe$ and is 
destroyed in the weak-field region. It must be stressed that the mentioned
above fit is made essentially by hand and without any error
estimation. However, it can serve as demonstration of effective
suppression of interlayer interaction in the high field region.\\ 

\section{Summary.}
In this paper we considered HTSC assuming weak interlayer nearest
neighboring coupling. In the region of the critical fluctuations, close to
the mean--field transition line $H_{c2}(T)$ the order parameter can be 
taken
as a linear combination of the electron eigenfunctions of the lowest
Landau level. This approximation together with the assumption that the
Abrikosov geometric factor only weakly depends on the actual vortex
configuration allows to reduce the problem to the simpler one, namely to
the saddle point equation (\ref{sp2.eq}). This equation can be
solved perturbatively. We calculated the magnetization of
the superconducting sample to the second order in the effective coupling
constant $\mu(T,H)$. If the coupling is sufficiently weak, one can
consider
the
first order correction only. It turns out that this correction does not
modify the scaling properties of the sample, leading to the trivial
renormalization of the critical temperature (see eq. (\ref{tshift.eq})).
Account of the second order
correction leads to violation of the 2D scaling and destroys
the magnetization crossing point in the low field region. 
At sufficiently high values of the applied field the interlayer interaction is
effectively suppressed. This leads to recovering of 2D scaling with a
well pronounced crossing point.\\

\section*{Acknowledgments}

The author would like to thank Sergey A. Gredeskul for helpful
discussions.\\

The author gratefully acknowledges the MINERVA foundation for the
financial support.

\end{document}